\documentstyle[12pt,titlepage]{article}
\begin{document}
\begin{titlepage}
\parindent 0pt
{\Large\bf
Universality in the Three-Dimensional Hard-Sphere Lattice Gas
}

\bigskip

{\large\bf Atsushi Yamagata}

{\it
Department of Physics, Tokyo Metropolitan University,
Minami-ohsawa 1-1, Hachioji-shi, Tokyo 192-03, Japan
}

\bigskip




\bigskip

{\bf Abstract.}

We perform Monte Carlo simulations of the hard-sphere lattice gas
on the body-centred cubic lattice with nearest neighbour exclusion.
We get the critical exponents,
$\beta/\nu = 0.311(8)$ and $\gamma/\nu = 2.38(2)$,
where $\beta$, $\gamma$, and $\nu$ are the critical exponents of
the staggered density, the staggered compressibility, and
the correlation length, respectively.
The values of the hard-sphere lattice gas on the simple cubic lattice
agree with them but those of the three-dimensional Ising model do not.
This supports that the hard-sphere lattice gas does not
fall into the Ising universality class in three dimensions.
\end{titlepage}

In this letter we study the hard-sphere lattice gas
whose atoms interact with infinite repulsion
of nearest neighbour pairs.
The grand partition function is
\begin{equation}
{\mit \Xi_{V}(z)}
=
\sum_{N} z^{N}\,Z_{V}(N),
\label{eqn:gpf}
\end{equation}
where $z$ is an activity and
$Z_{V}(N)$ is the number of configurations in which
there are $N$ atoms in the lattice of $V$ sites.
There are many studies on:
the square lattice~%
\cite{%
Domb58,%
Temperley59,%
Burley60,%
GauntFisher65,%
Runnels65,%
RunnelsCombs66,%
ReeChesnut66,%
Baxteretal80,%
Racz80,%
WoodGoldfinch80,%
Meirovitch83,%
KamieniarzBlote93a,%
KamieniarzBlote93b%
},
the triangular lattice~%
\cite{%
Burley60,%
RunnelsCombs66,%
Racz80,%
WoodGoldfinch80,%
Gaunt67,%
Baxter80,%
BaxterPearce82%
},
the honeycomb lattice~%
\cite{%
Runnelsetal67%
},
the simple cubic lattice~%
\cite{%
Domb58,%
Burley60,%
Racz80,%
Gaunt67,%
Yamagata95a,%
Yamagata95b%
},
the body-centred cubic lattice~%
\cite{%
Burley60,%
Racz80,%
Gaunt67,%
Yamagata95c%
},
and the face-centred cubic lattice~%
\cite{%
Burley60%
}.

On the simple cubic lattice
the author has estimated the critical exponents,
$\beta/\nu = 0.313(9)$ and $\gamma/\nu = 2.37(2)$,
where $\beta$, $\gamma$, and $\nu$ are the critical exponents
of the staggered density, the staggered compressibility, and
the correlation length, respectively~%
\cite{Yamagata95a}.
The corresponding values of the Ising model are
$\beta/\nu = 0.518(7)$ and $\gamma/\nu = 1.9828(57)$~%
\cite{%
FerrenbergLandau91%
}.
It does not seem that the hard-sphere lattice gas falls into
the Ising universality class in three dimensions.
The purpose of this letter is to obtain another piece of evidence
of that.
We estimate the critical exponents of the hard-sphere lattice gas
on the body-centred cubic lattice.

We carry out Monte Carlo simulations~%
\cite{%
Binder79,BinderStauffer87%
}
of the hard-sphere lattice gas~(\ref{eqn:gpf})
on the body-centred cubic lattice
of $V$ sites, where $V = 2 \times L \times L \times L$
($L$ = $2 \times n$, $n$ = 2, 3, $\ldots$ , 30),
under fully periodic boundary conditions~%
\cite{%
Meirovitch83,%
Yamagata95a%
}.
We measure the staggered density,
\begin{equation}
m^{\dagger \prime}
=
\overline{
\langle R^{2} \rangle^{1/2}
},
\label{eqn:rmsdensi}
\end{equation}
and the staggered compressibility,
\begin{equation}
\chi^{\dagger \prime}
=
V
\overline{
\langle R^{2} \rangle
}
/ 4,
\label{eqn:rmscompr}
\end{equation}
where $R=2\,(N_{\rm A}-N_{\rm B})/V$ and $N_{\rm A}$ ($N_{\rm B}$)
is the number of the atoms in the A (B)-sublattice.
Each run is divided into ten or twelve blocks.
$\langle \cdots \rangle$ is an expectation in a block and
$\overline{\cdots}$ is one over blocks.
All the simulations are done at the critical activity,
$z_{\rm c} = 0.7223$,~%
\cite{%
Yamagata95c%
}
over $12 \times 10^{5}$ Monte Carlo steps per site (MCS/site) or
$10 \times 10^{5}$ MCS/site
after discarding $5 \times 10^{4}$ MCS/site to attain equilibrium.
We have checked that simulations from the ground state configuration
(The atoms occupy all the sites of one sublattice and
the other is vacant.)
and no atom one gave consistent results.
The pseudorandom numbers are generated by the Tausworthe method~%
\cite{%
ItoKanada88,%
ItoKanada90%
}.

We estimate a critical exponent and an amplitude
by using the finite-size scaling~%
\cite{Fisher70,Barber83,Privman90}.
For a physical quantitiy, $O$,
we use the nonlinear chi-square fitting~%
\cite{%
Pressetal92%
}
with a function of $L$, $O(L)=A\,L^{p}$,
where $p$ and $A$ are fitting parameters.
$p=-\beta/\nu$ when $O$ = $m^{\dagger \prime}$.
$p=\gamma/\nu$ when $O$ = $\chi^{\dagger \prime}$.
We calculate the chi-square per degrees of freedom,
$\chi^{2}/{\rm DOF}$,
and the goodness of fit, $Q$,~%
\cite{%
Pressetal92%
}
for the data set of the sizes:
$L_{\rm min}, L_{\rm min}+2, \ldots , L_{\rm max}-2$, and
$L_{\rm max}$.
The values of $L_{\rm min}$ and $L_{\rm max}$ are selected so that
the difference between $\chi^{2}/{\rm DOF}$ and 1 is the smallest.

We get the results as follows.
For $m^{\dagger \prime}$ defined by (\ref{eqn:rmsdensi}),
$\beta/\nu$ = 0.311(8),
$A$ = 0.38(1),
$\chi^{2}/{\rm DOF} = 0.94$, and
$Q = 0.50$.
For $\chi^{\dagger \prime}$ defined by (\ref{eqn:rmscompr}),
$\gamma/\nu$ = 2.38(2),
$A$ = 0.037(2),
$\chi^{2}/{\rm DOF} = 0.93$, and
$Q = 0.50$.
$L_{\rm min} = 26 $ and $L_{\rm max} = 48$
for $m^{\dagger \prime}$ and $\chi^{\dagger \prime}$.
Figure~1 shows the size dependence of $m^{\dagger \prime}$
at $z = z_{\rm c}$.
The solid line indicates $0.38 L^{-0.311}$.
Figure~2 shows the size dependence of $\chi^{\dagger \prime}$
at $z = z_{\rm c}$.
The solid line indicates $0.037 L^{2.38}$.

The values of the critical exponents are consistent with
those of the hard-sphere lattice gas on the simple cubic lattice,
$\beta/\nu$ = 0.313(9) and $\gamma/\nu$ = 2.37(2)~%
\cite{%
Yamagata95b%
}.
They are different from the corresponding values of
the Ising model,
$\beta/\nu$ = 0.518(7) and $\gamma/\nu$ = 1.9828(57)~%
\cite{%
FerrenbergLandau91%
}.

In conclusion,
we have obtained strong evidence that
the hard-sphere lattice gas does not fall into
the Ising universality class in three dimensions.
This is an interesting result
since the models on the square lattice are
in the same universality class~%
\cite{%
Meirovitch83,%
KamieniarzBlote93a,%
KamieniarzBlote93b%
}.

\section*{Acknowledgements}
The author would like to thank Dr. Katsumi Kasono
for useful discussions and critical reading of the manuscript.
We have carried out the simulations on the HITAC S-3600/120 computer
under the Institute of Statistical Mathematics Cooperative Research
Program (95-ISM$\cdot$CRP-37) and
on the personal computer with the Pentium/120MHz CPU and
the Linux 1.2.8 operating system
(Slackware-2.3.0 + JE-0.9.6$\beta$2).
This study was supported by a Grant-in-Aid for Scientific Research
from the Ministry of Education, Science and Culture, Japan.


\clearpage
\section*{Figure captions}
\begin{description}
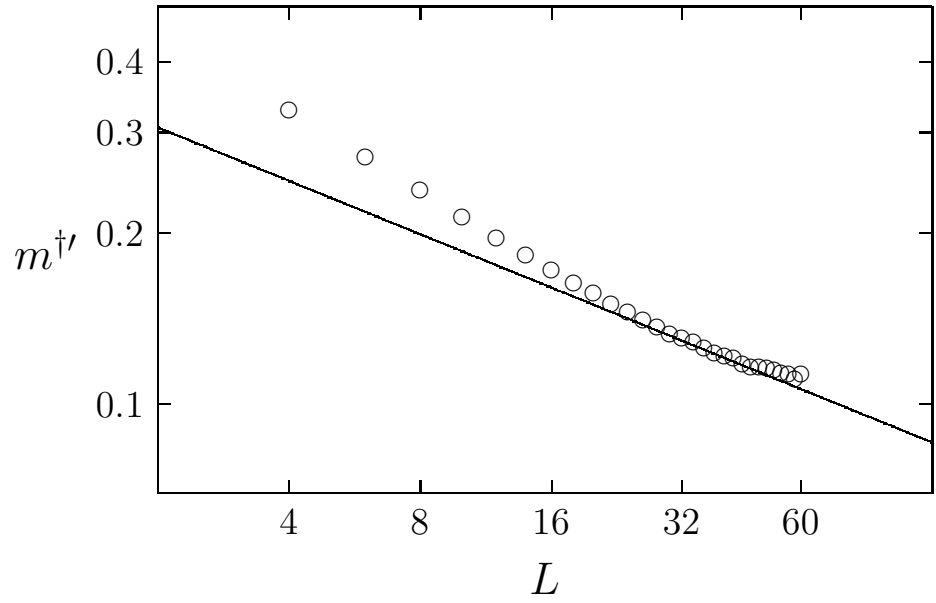
\item[Figure 1] Size dependence of the staggered density,
$m^{\dagger \prime}$, defined by (\ref{eqn:rmsdensi})
at $z = z_{\rm c}$.
The solid line indicates $0.38 L^{-0.311}$.
Errors are less than the symbol size.

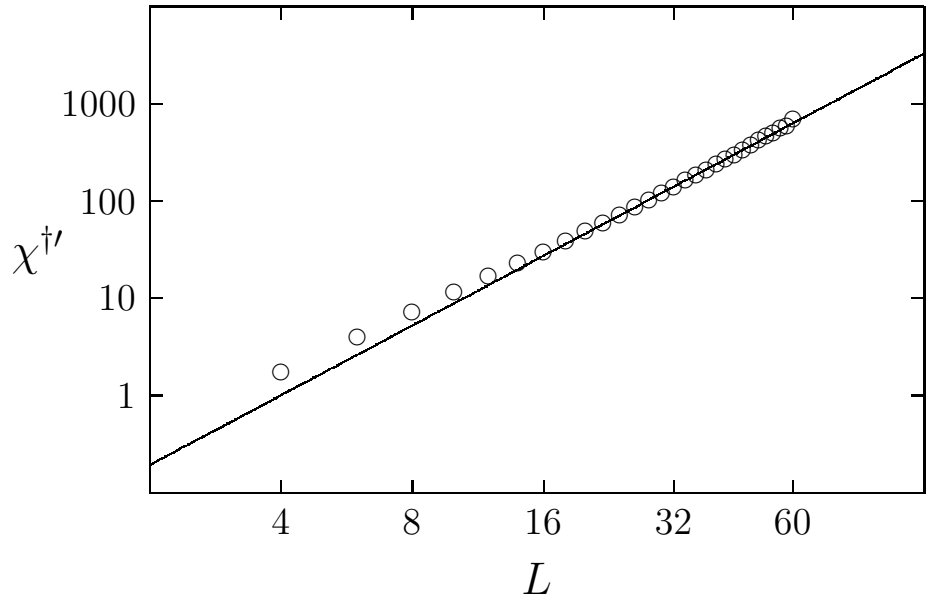
\item[Figure 2] Size dependence of the staggered compressibility,
$\chi^{\dagger \prime}$, defined by~(\ref{eqn:rmscompr})
at $z = z_{\rm c}$.
The solid line indicates $0.037 L^{2.38}$.
Errors are less than the symbol size.
\end{description}

\clearpage
\begin{figure}
\begin{center}
\setlength{\unitlength}{0.240900pt}
\ifx\plotpoint\undefined\newsavebox{\plotpoint}\fi
\sbox{\plotpoint}{\rule[-0.200pt]{0.400pt}{0.400pt}}%
\begin{picture}(1500,900)(0,0)
\font\gnuplot=cmr10 at 10pt
\gnuplot
\sbox{\plotpoint}{\rule[-0.200pt]{0.400pt}{0.400pt}}%
\put(220.0,252.0){\rule[-0.200pt]{4.818pt}{0.400pt}}
\put(198,252){\makebox(0,0)[r]{{\large $0.1$}}}
\put(1416.0,252.0){\rule[-0.200pt]{4.818pt}{0.400pt}}
\put(220.0,521.0){\rule[-0.200pt]{4.818pt}{0.400pt}}
\put(198,521){\makebox(0,0)[r]{{\large $0.2$}}}
\put(1416.0,521.0){\rule[-0.200pt]{4.818pt}{0.400pt}}
\put(220.0,679.0){\rule[-0.200pt]{4.818pt}{0.400pt}}
\put(198,679){\makebox(0,0)[r]{{\large $0.3$}}}
\put(1416.0,679.0){\rule[-0.200pt]{4.818pt}{0.400pt}}
\put(220.0,790.0){\rule[-0.200pt]{4.818pt}{0.400pt}}
\put(198,790){\makebox(0,0)[r]{{\large $0.4$}}}
\put(1416.0,790.0){\rule[-0.200pt]{4.818pt}{0.400pt}}
\put(426.0,113.0){\rule[-0.200pt]{0.400pt}{4.818pt}}
\put(426,68){\makebox(0,0){{\large $4$}}}
\put(426.0,857.0){\rule[-0.200pt]{0.400pt}{4.818pt}}
\put(632.0,113.0){\rule[-0.200pt]{0.400pt}{4.818pt}}
\put(632,68){\makebox(0,0){{\large $8$}}}
\put(632.0,857.0){\rule[-0.200pt]{0.400pt}{4.818pt}}
\put(838.0,113.0){\rule[-0.200pt]{0.400pt}{4.818pt}}
\put(838,68){\makebox(0,0){{\large $16$}}}
\put(838.0,857.0){\rule[-0.200pt]{0.400pt}{4.818pt}}
\put(1043.0,113.0){\rule[-0.200pt]{0.400pt}{4.818pt}}
\put(1043,68){\makebox(0,0){{\large $32$}}}
\put(1043.0,857.0){\rule[-0.200pt]{0.400pt}{4.818pt}}
\put(1230.0,113.0){\rule[-0.200pt]{0.400pt}{4.818pt}}
\put(1230,68){\makebox(0,0){{\large $60$}}}
\put(1230.0,857.0){\rule[-0.200pt]{0.400pt}{4.818pt}}
\put(220.0,113.0){\rule[-0.200pt]{292.934pt}{0.400pt}}
\put(1436.0,113.0){\rule[-0.200pt]{0.400pt}{184.048pt}}
\put(220.0,877.0){\rule[-0.200pt]{292.934pt}{0.400pt}}
\put(45,495){\makebox(0,0){{\Large $m^{\dagger \prime}$}}}
\put(828,-22){\makebox(0,0){{\Large $L$}}}
\put(818,-373){\makebox(0,0){{\large Figure 1}}}
\put(818,-485){\makebox(0,0){{\large Atsushi Yamagata}}}
\put(220.0,113.0){\rule[-0.200pt]{0.400pt}{184.048pt}}
\put(426,714){\circle{24}}
\put(546,640){\circle{24}}
\put(632,588){\circle{24}}
\put(698,547){\circle{24}}
\put(752,514){\circle{24}}
\put(798,487){\circle{24}}
\put(838,463){\circle{24}}
\put(873,443){\circle{24}}
\put(904,427){\circle{24}}
\put(932,409){\circle{24}}
\put(958,397){\circle{24}}
\put(982,385){\circle{24}}
\put(1004,373){\circle{24}}
\put(1024,363){\circle{24}}
\put(1043,356){\circle{24}}
\put(1061,350){\circle{24}}
\put(1078,341){\circle{24}}
\put(1094,333){\circle{24}}
\put(1110,328){\circle{24}}
\put(1124,325){\circle{24}}
\put(1138,315){\circle{24}}
\put(1151,311){\circle{24}}
\put(1164,310){\circle{24}}
\put(1176,309){\circle{24}}
\put(1188,306){\circle{24}}
\put(1199,301){\circle{24}}
\put(1210,299){\circle{24}}
\put(1220,292){\circle{24}}
\put(1230,300){\circle{24}}
\put(220,687){\usebox{\plotpoint}}
\multiput(220.00,685.93)(1.267,-0.477){7}{\rule{1.060pt}{0.115pt}}
\multiput(220.00,686.17)(9.800,-5.000){2}{\rule{0.530pt}{0.400pt}}
\multiput(232.00,680.93)(1.378,-0.477){7}{\rule{1.140pt}{0.115pt}}
\multiput(232.00,681.17)(10.634,-5.000){2}{\rule{0.570pt}{0.400pt}}
\multiput(245.00,675.93)(1.267,-0.477){7}{\rule{1.060pt}{0.115pt}}
\multiput(245.00,676.17)(9.800,-5.000){2}{\rule{0.530pt}{0.400pt}}
\multiput(257.00,670.93)(1.267,-0.477){7}{\rule{1.060pt}{0.115pt}}
\multiput(257.00,671.17)(9.800,-5.000){2}{\rule{0.530pt}{0.400pt}}
\multiput(269.00,665.93)(1.267,-0.477){7}{\rule{1.060pt}{0.115pt}}
\multiput(269.00,666.17)(9.800,-5.000){2}{\rule{0.530pt}{0.400pt}}
\multiput(281.00,660.93)(1.378,-0.477){7}{\rule{1.140pt}{0.115pt}}
\multiput(281.00,661.17)(10.634,-5.000){2}{\rule{0.570pt}{0.400pt}}
\multiput(294.00,655.93)(1.267,-0.477){7}{\rule{1.060pt}{0.115pt}}
\multiput(294.00,656.17)(9.800,-5.000){2}{\rule{0.530pt}{0.400pt}}
\multiput(306.00,650.93)(1.267,-0.477){7}{\rule{1.060pt}{0.115pt}}
\multiput(306.00,651.17)(9.800,-5.000){2}{\rule{0.530pt}{0.400pt}}
\multiput(318.00,645.93)(1.378,-0.477){7}{\rule{1.140pt}{0.115pt}}
\multiput(318.00,646.17)(10.634,-5.000){2}{\rule{0.570pt}{0.400pt}}
\multiput(331.00,640.93)(1.267,-0.477){7}{\rule{1.060pt}{0.115pt}}
\multiput(331.00,641.17)(9.800,-5.000){2}{\rule{0.530pt}{0.400pt}}
\multiput(343.00,635.93)(1.267,-0.477){7}{\rule{1.060pt}{0.115pt}}
\multiput(343.00,636.17)(9.800,-5.000){2}{\rule{0.530pt}{0.400pt}}
\multiput(355.00,630.93)(1.267,-0.477){7}{\rule{1.060pt}{0.115pt}}
\multiput(355.00,631.17)(9.800,-5.000){2}{\rule{0.530pt}{0.400pt}}
\multiput(367.00,625.93)(1.378,-0.477){7}{\rule{1.140pt}{0.115pt}}
\multiput(367.00,626.17)(10.634,-5.000){2}{\rule{0.570pt}{0.400pt}}
\multiput(380.00,620.93)(1.267,-0.477){7}{\rule{1.060pt}{0.115pt}}
\multiput(380.00,621.17)(9.800,-5.000){2}{\rule{0.530pt}{0.400pt}}
\multiput(392.00,615.93)(1.267,-0.477){7}{\rule{1.060pt}{0.115pt}}
\multiput(392.00,616.17)(9.800,-5.000){2}{\rule{0.530pt}{0.400pt}}
\multiput(404.00,610.93)(1.378,-0.477){7}{\rule{1.140pt}{0.115pt}}
\multiput(404.00,611.17)(10.634,-5.000){2}{\rule{0.570pt}{0.400pt}}
\multiput(417.00,605.93)(1.267,-0.477){7}{\rule{1.060pt}{0.115pt}}
\multiput(417.00,606.17)(9.800,-5.000){2}{\rule{0.530pt}{0.400pt}}
\multiput(429.00,600.93)(1.267,-0.477){7}{\rule{1.060pt}{0.115pt}}
\multiput(429.00,601.17)(9.800,-5.000){2}{\rule{0.530pt}{0.400pt}}
\multiput(441.00,595.93)(1.267,-0.477){7}{\rule{1.060pt}{0.115pt}}
\multiput(441.00,596.17)(9.800,-5.000){2}{\rule{0.530pt}{0.400pt}}
\multiput(453.00,590.93)(1.378,-0.477){7}{\rule{1.140pt}{0.115pt}}
\multiput(453.00,591.17)(10.634,-5.000){2}{\rule{0.570pt}{0.400pt}}
\multiput(466.00,585.93)(1.267,-0.477){7}{\rule{1.060pt}{0.115pt}}
\multiput(466.00,586.17)(9.800,-5.000){2}{\rule{0.530pt}{0.400pt}}
\multiput(478.00,580.93)(1.267,-0.477){7}{\rule{1.060pt}{0.115pt}}
\multiput(478.00,581.17)(9.800,-5.000){2}{\rule{0.530pt}{0.400pt}}
\multiput(490.00,575.93)(1.378,-0.477){7}{\rule{1.140pt}{0.115pt}}
\multiput(490.00,576.17)(10.634,-5.000){2}{\rule{0.570pt}{0.400pt}}
\multiput(503.00,570.93)(1.267,-0.477){7}{\rule{1.060pt}{0.115pt}}
\multiput(503.00,571.17)(9.800,-5.000){2}{\rule{0.530pt}{0.400pt}}
\multiput(515.00,565.93)(1.267,-0.477){7}{\rule{1.060pt}{0.115pt}}
\multiput(515.00,566.17)(9.800,-5.000){2}{\rule{0.530pt}{0.400pt}}
\multiput(527.00,560.93)(1.267,-0.477){7}{\rule{1.060pt}{0.115pt}}
\multiput(527.00,561.17)(9.800,-5.000){2}{\rule{0.530pt}{0.400pt}}
\multiput(539.00,555.93)(1.378,-0.477){7}{\rule{1.140pt}{0.115pt}}
\multiput(539.00,556.17)(10.634,-5.000){2}{\rule{0.570pt}{0.400pt}}
\multiput(552.00,550.93)(1.267,-0.477){7}{\rule{1.060pt}{0.115pt}}
\multiput(552.00,551.17)(9.800,-5.000){2}{\rule{0.530pt}{0.400pt}}
\multiput(564.00,545.93)(1.267,-0.477){7}{\rule{1.060pt}{0.115pt}}
\multiput(564.00,546.17)(9.800,-5.000){2}{\rule{0.530pt}{0.400pt}}
\multiput(576.00,540.93)(1.267,-0.477){7}{\rule{1.060pt}{0.115pt}}
\multiput(576.00,541.17)(9.800,-5.000){2}{\rule{0.530pt}{0.400pt}}
\multiput(588.00,535.93)(1.378,-0.477){7}{\rule{1.140pt}{0.115pt}}
\multiput(588.00,536.17)(10.634,-5.000){2}{\rule{0.570pt}{0.400pt}}
\multiput(601.00,530.93)(1.267,-0.477){7}{\rule{1.060pt}{0.115pt}}
\multiput(601.00,531.17)(9.800,-5.000){2}{\rule{0.530pt}{0.400pt}}
\multiput(613.00,525.93)(1.267,-0.477){7}{\rule{1.060pt}{0.115pt}}
\multiput(613.00,526.17)(9.800,-5.000){2}{\rule{0.530pt}{0.400pt}}
\multiput(625.00,520.93)(1.378,-0.477){7}{\rule{1.140pt}{0.115pt}}
\multiput(625.00,521.17)(10.634,-5.000){2}{\rule{0.570pt}{0.400pt}}
\multiput(638.00,515.93)(1.267,-0.477){7}{\rule{1.060pt}{0.115pt}}
\multiput(638.00,516.17)(9.800,-5.000){2}{\rule{0.530pt}{0.400pt}}
\multiput(650.00,510.93)(1.267,-0.477){7}{\rule{1.060pt}{0.115pt}}
\multiput(650.00,511.17)(9.800,-5.000){2}{\rule{0.530pt}{0.400pt}}
\multiput(662.00,505.93)(1.267,-0.477){7}{\rule{1.060pt}{0.115pt}}
\multiput(662.00,506.17)(9.800,-5.000){2}{\rule{0.530pt}{0.400pt}}
\multiput(674.00,500.93)(1.378,-0.477){7}{\rule{1.140pt}{0.115pt}}
\multiput(674.00,501.17)(10.634,-5.000){2}{\rule{0.570pt}{0.400pt}}
\multiput(687.00,495.93)(1.267,-0.477){7}{\rule{1.060pt}{0.115pt}}
\multiput(687.00,496.17)(9.800,-5.000){2}{\rule{0.530pt}{0.400pt}}
\multiput(699.00,490.93)(1.267,-0.477){7}{\rule{1.060pt}{0.115pt}}
\multiput(699.00,491.17)(9.800,-5.000){2}{\rule{0.530pt}{0.400pt}}
\multiput(711.00,485.93)(1.378,-0.477){7}{\rule{1.140pt}{0.115pt}}
\multiput(711.00,486.17)(10.634,-5.000){2}{\rule{0.570pt}{0.400pt}}
\multiput(724.00,480.93)(1.267,-0.477){7}{\rule{1.060pt}{0.115pt}}
\multiput(724.00,481.17)(9.800,-5.000){2}{\rule{0.530pt}{0.400pt}}
\multiput(736.00,475.93)(1.267,-0.477){7}{\rule{1.060pt}{0.115pt}}
\multiput(736.00,476.17)(9.800,-5.000){2}{\rule{0.530pt}{0.400pt}}
\multiput(748.00,470.93)(1.267,-0.477){7}{\rule{1.060pt}{0.115pt}}
\multiput(748.00,471.17)(9.800,-5.000){2}{\rule{0.530pt}{0.400pt}}
\multiput(760.00,465.93)(1.378,-0.477){7}{\rule{1.140pt}{0.115pt}}
\multiput(760.00,466.17)(10.634,-5.000){2}{\rule{0.570pt}{0.400pt}}
\multiput(773.00,460.93)(1.267,-0.477){7}{\rule{1.060pt}{0.115pt}}
\multiput(773.00,461.17)(9.800,-5.000){2}{\rule{0.530pt}{0.400pt}}
\multiput(785.00,455.93)(1.267,-0.477){7}{\rule{1.060pt}{0.115pt}}
\multiput(785.00,456.17)(9.800,-5.000){2}{\rule{0.530pt}{0.400pt}}
\multiput(797.00,450.93)(1.378,-0.477){7}{\rule{1.140pt}{0.115pt}}
\multiput(797.00,451.17)(10.634,-5.000){2}{\rule{0.570pt}{0.400pt}}
\multiput(810.00,445.93)(1.267,-0.477){7}{\rule{1.060pt}{0.115pt}}
\multiput(810.00,446.17)(9.800,-5.000){2}{\rule{0.530pt}{0.400pt}}
\multiput(822.00,440.93)(1.267,-0.477){7}{\rule{1.060pt}{0.115pt}}
\multiput(822.00,441.17)(9.800,-5.000){2}{\rule{0.530pt}{0.400pt}}
\multiput(834.00,435.93)(1.267,-0.477){7}{\rule{1.060pt}{0.115pt}}
\multiput(834.00,436.17)(9.800,-5.000){2}{\rule{0.530pt}{0.400pt}}
\multiput(846.00,430.93)(1.378,-0.477){7}{\rule{1.140pt}{0.115pt}}
\multiput(846.00,431.17)(10.634,-5.000){2}{\rule{0.570pt}{0.400pt}}
\multiput(859.00,425.93)(1.267,-0.477){7}{\rule{1.060pt}{0.115pt}}
\multiput(859.00,426.17)(9.800,-5.000){2}{\rule{0.530pt}{0.400pt}}
\multiput(871.00,420.93)(1.267,-0.477){7}{\rule{1.060pt}{0.115pt}}
\multiput(871.00,421.17)(9.800,-5.000){2}{\rule{0.530pt}{0.400pt}}
\multiput(883.00,415.93)(1.378,-0.477){7}{\rule{1.140pt}{0.115pt}}
\multiput(883.00,416.17)(10.634,-5.000){2}{\rule{0.570pt}{0.400pt}}
\multiput(896.00,410.93)(1.267,-0.477){7}{\rule{1.060pt}{0.115pt}}
\multiput(896.00,411.17)(9.800,-5.000){2}{\rule{0.530pt}{0.400pt}}
\multiput(908.00,405.93)(1.267,-0.477){7}{\rule{1.060pt}{0.115pt}}
\multiput(908.00,406.17)(9.800,-5.000){2}{\rule{0.530pt}{0.400pt}}
\multiput(920.00,400.93)(1.267,-0.477){7}{\rule{1.060pt}{0.115pt}}
\multiput(920.00,401.17)(9.800,-5.000){2}{\rule{0.530pt}{0.400pt}}
\multiput(932.00,395.93)(1.378,-0.477){7}{\rule{1.140pt}{0.115pt}}
\multiput(932.00,396.17)(10.634,-5.000){2}{\rule{0.570pt}{0.400pt}}
\multiput(945.00,390.93)(1.267,-0.477){7}{\rule{1.060pt}{0.115pt}}
\multiput(945.00,391.17)(9.800,-5.000){2}{\rule{0.530pt}{0.400pt}}
\multiput(957.00,385.93)(1.267,-0.477){7}{\rule{1.060pt}{0.115pt}}
\multiput(957.00,386.17)(9.800,-5.000){2}{\rule{0.530pt}{0.400pt}}
\multiput(969.00,380.93)(1.378,-0.477){7}{\rule{1.140pt}{0.115pt}}
\multiput(969.00,381.17)(10.634,-5.000){2}{\rule{0.570pt}{0.400pt}}
\multiput(982.00,375.93)(1.267,-0.477){7}{\rule{1.060pt}{0.115pt}}
\multiput(982.00,376.17)(9.800,-5.000){2}{\rule{0.530pt}{0.400pt}}
\multiput(994.00,370.93)(1.267,-0.477){7}{\rule{1.060pt}{0.115pt}}
\multiput(994.00,371.17)(9.800,-5.000){2}{\rule{0.530pt}{0.400pt}}
\multiput(1006.00,365.93)(1.267,-0.477){7}{\rule{1.060pt}{0.115pt}}
\multiput(1006.00,366.17)(9.800,-5.000){2}{\rule{0.530pt}{0.400pt}}
\multiput(1018.00,360.93)(1.378,-0.477){7}{\rule{1.140pt}{0.115pt}}
\multiput(1018.00,361.17)(10.634,-5.000){2}{\rule{0.570pt}{0.400pt}}
\multiput(1031.00,355.93)(1.267,-0.477){7}{\rule{1.060pt}{0.115pt}}
\multiput(1031.00,356.17)(9.800,-5.000){2}{\rule{0.530pt}{0.400pt}}
\multiput(1043.00,350.93)(1.267,-0.477){7}{\rule{1.060pt}{0.115pt}}
\multiput(1043.00,351.17)(9.800,-5.000){2}{\rule{0.530pt}{0.400pt}}
\multiput(1055.00,345.93)(1.378,-0.477){7}{\rule{1.140pt}{0.115pt}}
\multiput(1055.00,346.17)(10.634,-5.000){2}{\rule{0.570pt}{0.400pt}}
\multiput(1068.00,340.93)(1.267,-0.477){7}{\rule{1.060pt}{0.115pt}}
\multiput(1068.00,341.17)(9.800,-5.000){2}{\rule{0.530pt}{0.400pt}}
\multiput(1080.00,335.93)(1.267,-0.477){7}{\rule{1.060pt}{0.115pt}}
\multiput(1080.00,336.17)(9.800,-5.000){2}{\rule{0.530pt}{0.400pt}}
\multiput(1092.00,330.93)(1.267,-0.477){7}{\rule{1.060pt}{0.115pt}}
\multiput(1092.00,331.17)(9.800,-5.000){2}{\rule{0.530pt}{0.400pt}}
\multiput(1104.00,325.93)(1.378,-0.477){7}{\rule{1.140pt}{0.115pt}}
\multiput(1104.00,326.17)(10.634,-5.000){2}{\rule{0.570pt}{0.400pt}}
\multiput(1117.00,320.93)(1.267,-0.477){7}{\rule{1.060pt}{0.115pt}}
\multiput(1117.00,321.17)(9.800,-5.000){2}{\rule{0.530pt}{0.400pt}}
\multiput(1129.00,315.93)(1.267,-0.477){7}{\rule{1.060pt}{0.115pt}}
\multiput(1129.00,316.17)(9.800,-5.000){2}{\rule{0.530pt}{0.400pt}}
\multiput(1141.00,310.93)(1.267,-0.477){7}{\rule{1.060pt}{0.115pt}}
\multiput(1141.00,311.17)(9.800,-5.000){2}{\rule{0.530pt}{0.400pt}}
\multiput(1153.00,305.93)(1.378,-0.477){7}{\rule{1.140pt}{0.115pt}}
\multiput(1153.00,306.17)(10.634,-5.000){2}{\rule{0.570pt}{0.400pt}}
\multiput(1166.00,300.93)(1.267,-0.477){7}{\rule{1.060pt}{0.115pt}}
\multiput(1166.00,301.17)(9.800,-5.000){2}{\rule{0.530pt}{0.400pt}}
\multiput(1178.00,295.93)(1.267,-0.477){7}{\rule{1.060pt}{0.115pt}}
\multiput(1178.00,296.17)(9.800,-5.000){2}{\rule{0.530pt}{0.400pt}}
\multiput(1190.00,290.93)(1.378,-0.477){7}{\rule{1.140pt}{0.115pt}}
\multiput(1190.00,291.17)(10.634,-5.000){2}{\rule{0.570pt}{0.400pt}}
\multiput(1203.00,285.93)(1.267,-0.477){7}{\rule{1.060pt}{0.115pt}}
\multiput(1203.00,286.17)(9.800,-5.000){2}{\rule{0.530pt}{0.400pt}}
\multiput(1215.00,280.93)(1.267,-0.477){7}{\rule{1.060pt}{0.115pt}}
\multiput(1215.00,281.17)(9.800,-5.000){2}{\rule{0.530pt}{0.400pt}}
\multiput(1227.00,275.93)(1.267,-0.477){7}{\rule{1.060pt}{0.115pt}}
\multiput(1227.00,276.17)(9.800,-5.000){2}{\rule{0.530pt}{0.400pt}}
\multiput(1239.00,270.93)(1.378,-0.477){7}{\rule{1.140pt}{0.115pt}}
\multiput(1239.00,271.17)(10.634,-5.000){2}{\rule{0.570pt}{0.400pt}}
\multiput(1252.00,265.93)(1.267,-0.477){7}{\rule{1.060pt}{0.115pt}}
\multiput(1252.00,266.17)(9.800,-5.000){2}{\rule{0.530pt}{0.400pt}}
\multiput(1264.00,260.93)(1.267,-0.477){7}{\rule{1.060pt}{0.115pt}}
\multiput(1264.00,261.17)(9.800,-5.000){2}{\rule{0.530pt}{0.400pt}}
\multiput(1276.00,255.93)(1.378,-0.477){7}{\rule{1.140pt}{0.115pt}}
\multiput(1276.00,256.17)(10.634,-5.000){2}{\rule{0.570pt}{0.400pt}}
\multiput(1289.00,250.93)(1.267,-0.477){7}{\rule{1.060pt}{0.115pt}}
\multiput(1289.00,251.17)(9.800,-5.000){2}{\rule{0.530pt}{0.400pt}}
\multiput(1301.00,245.93)(1.267,-0.477){7}{\rule{1.060pt}{0.115pt}}
\multiput(1301.00,246.17)(9.800,-5.000){2}{\rule{0.530pt}{0.400pt}}
\multiput(1313.00,240.93)(1.267,-0.477){7}{\rule{1.060pt}{0.115pt}}
\multiput(1313.00,241.17)(9.800,-5.000){2}{\rule{0.530pt}{0.400pt}}
\multiput(1325.00,235.93)(1.378,-0.477){7}{\rule{1.140pt}{0.115pt}}
\multiput(1325.00,236.17)(10.634,-5.000){2}{\rule{0.570pt}{0.400pt}}
\multiput(1338.00,230.93)(1.267,-0.477){7}{\rule{1.060pt}{0.115pt}}
\multiput(1338.00,231.17)(9.800,-5.000){2}{\rule{0.530pt}{0.400pt}}
\multiput(1350.00,225.93)(1.267,-0.477){7}{\rule{1.060pt}{0.115pt}}
\multiput(1350.00,226.17)(9.800,-5.000){2}{\rule{0.530pt}{0.400pt}}
\multiput(1362.00,220.93)(1.378,-0.477){7}{\rule{1.140pt}{0.115pt}}
\multiput(1362.00,221.17)(10.634,-5.000){2}{\rule{0.570pt}{0.400pt}}
\multiput(1375.00,215.93)(1.267,-0.477){7}{\rule{1.060pt}{0.115pt}}
\multiput(1375.00,216.17)(9.800,-5.000){2}{\rule{0.530pt}{0.400pt}}
\multiput(1387.00,210.93)(1.267,-0.477){7}{\rule{1.060pt}{0.115pt}}
\multiput(1387.00,211.17)(9.800,-5.000){2}{\rule{0.530pt}{0.400pt}}
\multiput(1399.00,205.93)(1.267,-0.477){7}{\rule{1.060pt}{0.115pt}}
\multiput(1399.00,206.17)(9.800,-5.000){2}{\rule{0.530pt}{0.400pt}}
\multiput(1411.00,200.93)(1.378,-0.477){7}{\rule{1.140pt}{0.115pt}}
\multiput(1411.00,201.17)(10.634,-5.000){2}{\rule{0.570pt}{0.400pt}}
\multiput(1424.00,195.93)(1.267,-0.477){7}{\rule{1.060pt}{0.115pt}}
\multiput(1424.00,196.17)(9.800,-5.000){2}{\rule{0.530pt}{0.400pt}}
\end{picture}
\end{center}
\end{figure}

\clearpage
\begin{figure}
\begin{center}
\setlength{\unitlength}{0.240900pt}
\ifx\plotpoint\undefined\newsavebox{\plotpoint}\fi
\sbox{\plotpoint}{\rule[-0.200pt]{0.400pt}{0.400pt}}%
\begin{picture}(1500,900)(0,0)
\font\gnuplot=cmr10 at 10pt
\gnuplot
\sbox{\plotpoint}{\rule[-0.200pt]{0.400pt}{0.400pt}}%
\put(220.0,266.0){\rule[-0.200pt]{4.818pt}{0.400pt}}
\put(198,266){\makebox(0,0)[r]{{\large $1$}}}
\put(1416.0,266.0){\rule[-0.200pt]{4.818pt}{0.400pt}}
\put(220.0,419.0){\rule[-0.200pt]{4.818pt}{0.400pt}}
\put(198,419){\makebox(0,0)[r]{{\large $10$}}}
\put(1416.0,419.0){\rule[-0.200pt]{4.818pt}{0.400pt}}
\put(220.0,571.0){\rule[-0.200pt]{4.818pt}{0.400pt}}
\put(198,571){\makebox(0,0)[r]{{\large $100$}}}
\put(1416.0,571.0){\rule[-0.200pt]{4.818pt}{0.400pt}}
\put(220.0,724.0){\rule[-0.200pt]{4.818pt}{0.400pt}}
\put(198,724){\makebox(0,0)[r]{{\large $1000$}}}
\put(1416.0,724.0){\rule[-0.200pt]{4.818pt}{0.400pt}}
\put(426.0,113.0){\rule[-0.200pt]{0.400pt}{4.818pt}}
\put(426,68){\makebox(0,0){{\large $4$}}}
\put(426.0,857.0){\rule[-0.200pt]{0.400pt}{4.818pt}}
\put(632.0,113.0){\rule[-0.200pt]{0.400pt}{4.818pt}}
\put(632,68){\makebox(0,0){{\large $8$}}}
\put(632.0,857.0){\rule[-0.200pt]{0.400pt}{4.818pt}}
\put(838.0,113.0){\rule[-0.200pt]{0.400pt}{4.818pt}}
\put(838,68){\makebox(0,0){{\large $16$}}}
\put(838.0,857.0){\rule[-0.200pt]{0.400pt}{4.818pt}}
\put(1043.0,113.0){\rule[-0.200pt]{0.400pt}{4.818pt}}
\put(1043,68){\makebox(0,0){{\large $32$}}}
\put(1043.0,857.0){\rule[-0.200pt]{0.400pt}{4.818pt}}
\put(1230.0,113.0){\rule[-0.200pt]{0.400pt}{4.818pt}}
\put(1230,68){\makebox(0,0){{\large $60$}}}
\put(1230.0,857.0){\rule[-0.200pt]{0.400pt}{4.818pt}}
\put(220.0,113.0){\rule[-0.200pt]{292.934pt}{0.400pt}}
\put(1436.0,113.0){\rule[-0.200pt]{0.400pt}{184.048pt}}
\put(220.0,877.0){\rule[-0.200pt]{292.934pt}{0.400pt}}
\put(45,495){\makebox(0,0){{\Large $\chi^{\dagger \prime}$}}}
\put(828,-22){\makebox(0,0){{\Large $L$}}}
\put(818,-344){\makebox(0,0){{\large Figure 2}}}
\put(818,-497){\makebox(0,0){{\large Atsushi Yamagata}}}
\put(220.0,113.0){\rule[-0.200pt]{0.400pt}{184.048pt}}
\put(426,302){\circle{24}}
\put(546,358){\circle{24}}
\put(632,397){\circle{24}}
\put(698,428){\circle{24}}
\put(752,453){\circle{24}}
\put(798,474){\circle{24}}
\put(838,492){\circle{24}}
\put(873,509){\circle{24}}
\put(904,525){\circle{24}}
\put(932,537){\circle{24}}
\put(958,550){\circle{24}}
\put(982,562){\circle{24}}
\put(1004,573){\circle{24}}
\put(1024,584){\circle{24}}
\put(1043,594){\circle{24}}
\put(1061,604){\circle{24}}
\put(1078,612){\circle{24}}
\put(1094,620){\circle{24}}
\put(1110,629){\circle{24}}
\put(1124,637){\circle{24}}
\put(1138,643){\circle{24}}
\put(1151,651){\circle{24}}
\put(1164,659){\circle{24}}
\put(1176,667){\circle{24}}
\put(1188,673){\circle{24}}
\put(1199,679){\circle{24}}
\put(1210,686){\circle{24}}
\put(1220,690){\circle{24}}
\put(1230,700){\circle{24}}
\put(220,156){\usebox{\plotpoint}}
\multiput(220.00,156.59)(0.874,0.485){11}{\rule{0.786pt}{0.117pt}}
\multiput(220.00,155.17)(10.369,7.000){2}{\rule{0.393pt}{0.400pt}}
\multiput(232.00,163.59)(0.950,0.485){11}{\rule{0.843pt}{0.117pt}}
\multiput(232.00,162.17)(11.251,7.000){2}{\rule{0.421pt}{0.400pt}}
\multiput(245.00,170.59)(1.033,0.482){9}{\rule{0.900pt}{0.116pt}}
\multiput(245.00,169.17)(10.132,6.000){2}{\rule{0.450pt}{0.400pt}}
\multiput(257.00,176.59)(0.874,0.485){11}{\rule{0.786pt}{0.117pt}}
\multiput(257.00,175.17)(10.369,7.000){2}{\rule{0.393pt}{0.400pt}}
\multiput(269.00,183.59)(1.033,0.482){9}{\rule{0.900pt}{0.116pt}}
\multiput(269.00,182.17)(10.132,6.000){2}{\rule{0.450pt}{0.400pt}}
\multiput(281.00,189.59)(0.950,0.485){11}{\rule{0.843pt}{0.117pt}}
\multiput(281.00,188.17)(11.251,7.000){2}{\rule{0.421pt}{0.400pt}}
\multiput(294.00,196.59)(1.033,0.482){9}{\rule{0.900pt}{0.116pt}}
\multiput(294.00,195.17)(10.132,6.000){2}{\rule{0.450pt}{0.400pt}}
\multiput(306.00,202.59)(0.874,0.485){11}{\rule{0.786pt}{0.117pt}}
\multiput(306.00,201.17)(10.369,7.000){2}{\rule{0.393pt}{0.400pt}}
\multiput(318.00,209.59)(1.123,0.482){9}{\rule{0.967pt}{0.116pt}}
\multiput(318.00,208.17)(10.994,6.000){2}{\rule{0.483pt}{0.400pt}}
\multiput(331.00,215.59)(0.874,0.485){11}{\rule{0.786pt}{0.117pt}}
\multiput(331.00,214.17)(10.369,7.000){2}{\rule{0.393pt}{0.400pt}}
\multiput(343.00,222.59)(1.033,0.482){9}{\rule{0.900pt}{0.116pt}}
\multiput(343.00,221.17)(10.132,6.000){2}{\rule{0.450pt}{0.400pt}}
\multiput(355.00,228.59)(0.874,0.485){11}{\rule{0.786pt}{0.117pt}}
\multiput(355.00,227.17)(10.369,7.000){2}{\rule{0.393pt}{0.400pt}}
\multiput(367.00,235.59)(1.123,0.482){9}{\rule{0.967pt}{0.116pt}}
\multiput(367.00,234.17)(10.994,6.000){2}{\rule{0.483pt}{0.400pt}}
\multiput(380.00,241.59)(0.874,0.485){11}{\rule{0.786pt}{0.117pt}}
\multiput(380.00,240.17)(10.369,7.000){2}{\rule{0.393pt}{0.400pt}}
\multiput(392.00,248.59)(1.033,0.482){9}{\rule{0.900pt}{0.116pt}}
\multiput(392.00,247.17)(10.132,6.000){2}{\rule{0.450pt}{0.400pt}}
\multiput(404.00,254.59)(0.950,0.485){11}{\rule{0.843pt}{0.117pt}}
\multiput(404.00,253.17)(11.251,7.000){2}{\rule{0.421pt}{0.400pt}}
\multiput(417.00,261.59)(0.874,0.485){11}{\rule{0.786pt}{0.117pt}}
\multiput(417.00,260.17)(10.369,7.000){2}{\rule{0.393pt}{0.400pt}}
\multiput(429.00,268.59)(1.033,0.482){9}{\rule{0.900pt}{0.116pt}}
\multiput(429.00,267.17)(10.132,6.000){2}{\rule{0.450pt}{0.400pt}}
\multiput(441.00,274.59)(0.874,0.485){11}{\rule{0.786pt}{0.117pt}}
\multiput(441.00,273.17)(10.369,7.000){2}{\rule{0.393pt}{0.400pt}}
\multiput(453.00,281.59)(1.123,0.482){9}{\rule{0.967pt}{0.116pt}}
\multiput(453.00,280.17)(10.994,6.000){2}{\rule{0.483pt}{0.400pt}}
\multiput(466.00,287.59)(0.874,0.485){11}{\rule{0.786pt}{0.117pt}}
\multiput(466.00,286.17)(10.369,7.000){2}{\rule{0.393pt}{0.400pt}}
\multiput(478.00,294.59)(1.033,0.482){9}{\rule{0.900pt}{0.116pt}}
\multiput(478.00,293.17)(10.132,6.000){2}{\rule{0.450pt}{0.400pt}}
\multiput(490.00,300.59)(0.950,0.485){11}{\rule{0.843pt}{0.117pt}}
\multiput(490.00,299.17)(11.251,7.000){2}{\rule{0.421pt}{0.400pt}}
\multiput(503.00,307.59)(1.033,0.482){9}{\rule{0.900pt}{0.116pt}}
\multiput(503.00,306.17)(10.132,6.000){2}{\rule{0.450pt}{0.400pt}}
\multiput(515.00,313.59)(0.874,0.485){11}{\rule{0.786pt}{0.117pt}}
\multiput(515.00,312.17)(10.369,7.000){2}{\rule{0.393pt}{0.400pt}}
\multiput(527.00,320.59)(1.033,0.482){9}{\rule{0.900pt}{0.116pt}}
\multiput(527.00,319.17)(10.132,6.000){2}{\rule{0.450pt}{0.400pt}}
\multiput(539.00,326.59)(0.950,0.485){11}{\rule{0.843pt}{0.117pt}}
\multiput(539.00,325.17)(11.251,7.000){2}{\rule{0.421pt}{0.400pt}}
\multiput(552.00,333.59)(1.033,0.482){9}{\rule{0.900pt}{0.116pt}}
\multiput(552.00,332.17)(10.132,6.000){2}{\rule{0.450pt}{0.400pt}}
\multiput(564.00,339.59)(0.874,0.485){11}{\rule{0.786pt}{0.117pt}}
\multiput(564.00,338.17)(10.369,7.000){2}{\rule{0.393pt}{0.400pt}}
\multiput(576.00,346.59)(1.033,0.482){9}{\rule{0.900pt}{0.116pt}}
\multiput(576.00,345.17)(10.132,6.000){2}{\rule{0.450pt}{0.400pt}}
\multiput(588.00,352.59)(0.950,0.485){11}{\rule{0.843pt}{0.117pt}}
\multiput(588.00,351.17)(11.251,7.000){2}{\rule{0.421pt}{0.400pt}}
\multiput(601.00,359.59)(0.874,0.485){11}{\rule{0.786pt}{0.117pt}}
\multiput(601.00,358.17)(10.369,7.000){2}{\rule{0.393pt}{0.400pt}}
\multiput(613.00,366.59)(1.033,0.482){9}{\rule{0.900pt}{0.116pt}}
\multiput(613.00,365.17)(10.132,6.000){2}{\rule{0.450pt}{0.400pt}}
\multiput(625.00,372.59)(0.950,0.485){11}{\rule{0.843pt}{0.117pt}}
\multiput(625.00,371.17)(11.251,7.000){2}{\rule{0.421pt}{0.400pt}}
\multiput(638.00,379.59)(1.033,0.482){9}{\rule{0.900pt}{0.116pt}}
\multiput(638.00,378.17)(10.132,6.000){2}{\rule{0.450pt}{0.400pt}}
\multiput(650.00,385.59)(0.874,0.485){11}{\rule{0.786pt}{0.117pt}}
\multiput(650.00,384.17)(10.369,7.000){2}{\rule{0.393pt}{0.400pt}}
\multiput(662.00,392.59)(1.033,0.482){9}{\rule{0.900pt}{0.116pt}}
\multiput(662.00,391.17)(10.132,6.000){2}{\rule{0.450pt}{0.400pt}}
\multiput(674.00,398.59)(0.950,0.485){11}{\rule{0.843pt}{0.117pt}}
\multiput(674.00,397.17)(11.251,7.000){2}{\rule{0.421pt}{0.400pt}}
\multiput(687.00,405.59)(1.033,0.482){9}{\rule{0.900pt}{0.116pt}}
\multiput(687.00,404.17)(10.132,6.000){2}{\rule{0.450pt}{0.400pt}}
\multiput(699.00,411.59)(0.874,0.485){11}{\rule{0.786pt}{0.117pt}}
\multiput(699.00,410.17)(10.369,7.000){2}{\rule{0.393pt}{0.400pt}}
\multiput(711.00,418.59)(1.123,0.482){9}{\rule{0.967pt}{0.116pt}}
\multiput(711.00,417.17)(10.994,6.000){2}{\rule{0.483pt}{0.400pt}}
\multiput(724.00,424.59)(0.874,0.485){11}{\rule{0.786pt}{0.117pt}}
\multiput(724.00,423.17)(10.369,7.000){2}{\rule{0.393pt}{0.400pt}}
\multiput(736.00,431.59)(1.033,0.482){9}{\rule{0.900pt}{0.116pt}}
\multiput(736.00,430.17)(10.132,6.000){2}{\rule{0.450pt}{0.400pt}}
\multiput(748.00,437.59)(0.874,0.485){11}{\rule{0.786pt}{0.117pt}}
\multiput(748.00,436.17)(10.369,7.000){2}{\rule{0.393pt}{0.400pt}}
\multiput(760.00,444.59)(1.123,0.482){9}{\rule{0.967pt}{0.116pt}}
\multiput(760.00,443.17)(10.994,6.000){2}{\rule{0.483pt}{0.400pt}}
\multiput(773.00,450.59)(0.874,0.485){11}{\rule{0.786pt}{0.117pt}}
\multiput(773.00,449.17)(10.369,7.000){2}{\rule{0.393pt}{0.400pt}}
\multiput(785.00,457.59)(1.033,0.482){9}{\rule{0.900pt}{0.116pt}}
\multiput(785.00,456.17)(10.132,6.000){2}{\rule{0.450pt}{0.400pt}}
\multiput(797.00,463.59)(0.950,0.485){11}{\rule{0.843pt}{0.117pt}}
\multiput(797.00,462.17)(11.251,7.000){2}{\rule{0.421pt}{0.400pt}}
\multiput(810.00,470.59)(0.874,0.485){11}{\rule{0.786pt}{0.117pt}}
\multiput(810.00,469.17)(10.369,7.000){2}{\rule{0.393pt}{0.400pt}}
\multiput(822.00,477.59)(1.033,0.482){9}{\rule{0.900pt}{0.116pt}}
\multiput(822.00,476.17)(10.132,6.000){2}{\rule{0.450pt}{0.400pt}}
\multiput(834.00,483.59)(0.874,0.485){11}{\rule{0.786pt}{0.117pt}}
\multiput(834.00,482.17)(10.369,7.000){2}{\rule{0.393pt}{0.400pt}}
\multiput(846.00,490.59)(1.123,0.482){9}{\rule{0.967pt}{0.116pt}}
\multiput(846.00,489.17)(10.994,6.000){2}{\rule{0.483pt}{0.400pt}}
\multiput(859.00,496.59)(0.874,0.485){11}{\rule{0.786pt}{0.117pt}}
\multiput(859.00,495.17)(10.369,7.000){2}{\rule{0.393pt}{0.400pt}}
\multiput(871.00,503.59)(1.033,0.482){9}{\rule{0.900pt}{0.116pt}}
\multiput(871.00,502.17)(10.132,6.000){2}{\rule{0.450pt}{0.400pt}}
\multiput(883.00,509.59)(0.950,0.485){11}{\rule{0.843pt}{0.117pt}}
\multiput(883.00,508.17)(11.251,7.000){2}{\rule{0.421pt}{0.400pt}}
\multiput(896.00,516.59)(1.033,0.482){9}{\rule{0.900pt}{0.116pt}}
\multiput(896.00,515.17)(10.132,6.000){2}{\rule{0.450pt}{0.400pt}}
\multiput(908.00,522.59)(0.874,0.485){11}{\rule{0.786pt}{0.117pt}}
\multiput(908.00,521.17)(10.369,7.000){2}{\rule{0.393pt}{0.400pt}}
\multiput(920.00,529.59)(1.033,0.482){9}{\rule{0.900pt}{0.116pt}}
\multiput(920.00,528.17)(10.132,6.000){2}{\rule{0.450pt}{0.400pt}}
\multiput(932.00,535.59)(0.950,0.485){11}{\rule{0.843pt}{0.117pt}}
\multiput(932.00,534.17)(11.251,7.000){2}{\rule{0.421pt}{0.400pt}}
\multiput(945.00,542.59)(1.033,0.482){9}{\rule{0.900pt}{0.116pt}}
\multiput(945.00,541.17)(10.132,6.000){2}{\rule{0.450pt}{0.400pt}}
\multiput(957.00,548.59)(0.874,0.485){11}{\rule{0.786pt}{0.117pt}}
\multiput(957.00,547.17)(10.369,7.000){2}{\rule{0.393pt}{0.400pt}}
\multiput(969.00,555.59)(1.123,0.482){9}{\rule{0.967pt}{0.116pt}}
\multiput(969.00,554.17)(10.994,6.000){2}{\rule{0.483pt}{0.400pt}}
\multiput(982.00,561.59)(0.874,0.485){11}{\rule{0.786pt}{0.117pt}}
\multiput(982.00,560.17)(10.369,7.000){2}{\rule{0.393pt}{0.400pt}}
\multiput(994.00,568.59)(0.874,0.485){11}{\rule{0.786pt}{0.117pt}}
\multiput(994.00,567.17)(10.369,7.000){2}{\rule{0.393pt}{0.400pt}}
\multiput(1006.00,575.59)(1.033,0.482){9}{\rule{0.900pt}{0.116pt}}
\multiput(1006.00,574.17)(10.132,6.000){2}{\rule{0.450pt}{0.400pt}}
\multiput(1018.00,581.59)(0.950,0.485){11}{\rule{0.843pt}{0.117pt}}
\multiput(1018.00,580.17)(11.251,7.000){2}{\rule{0.421pt}{0.400pt}}
\multiput(1031.00,588.59)(1.033,0.482){9}{\rule{0.900pt}{0.116pt}}
\multiput(1031.00,587.17)(10.132,6.000){2}{\rule{0.450pt}{0.400pt}}
\multiput(1043.00,594.59)(0.874,0.485){11}{\rule{0.786pt}{0.117pt}}
\multiput(1043.00,593.17)(10.369,7.000){2}{\rule{0.393pt}{0.400pt}}
\multiput(1055.00,601.59)(1.123,0.482){9}{\rule{0.967pt}{0.116pt}}
\multiput(1055.00,600.17)(10.994,6.000){2}{\rule{0.483pt}{0.400pt}}
\multiput(1068.00,607.59)(0.874,0.485){11}{\rule{0.786pt}{0.117pt}}
\multiput(1068.00,606.17)(10.369,7.000){2}{\rule{0.393pt}{0.400pt}}
\multiput(1080.00,614.59)(1.033,0.482){9}{\rule{0.900pt}{0.116pt}}
\multiput(1080.00,613.17)(10.132,6.000){2}{\rule{0.450pt}{0.400pt}}
\multiput(1092.00,620.59)(0.874,0.485){11}{\rule{0.786pt}{0.117pt}}
\multiput(1092.00,619.17)(10.369,7.000){2}{\rule{0.393pt}{0.400pt}}
\multiput(1104.00,627.59)(1.123,0.482){9}{\rule{0.967pt}{0.116pt}}
\multiput(1104.00,626.17)(10.994,6.000){2}{\rule{0.483pt}{0.400pt}}
\multiput(1117.00,633.59)(0.874,0.485){11}{\rule{0.786pt}{0.117pt}}
\multiput(1117.00,632.17)(10.369,7.000){2}{\rule{0.393pt}{0.400pt}}
\multiput(1129.00,640.59)(1.033,0.482){9}{\rule{0.900pt}{0.116pt}}
\multiput(1129.00,639.17)(10.132,6.000){2}{\rule{0.450pt}{0.400pt}}
\multiput(1141.00,646.59)(0.874,0.485){11}{\rule{0.786pt}{0.117pt}}
\multiput(1141.00,645.17)(10.369,7.000){2}{\rule{0.393pt}{0.400pt}}
\multiput(1153.00,653.59)(1.123,0.482){9}{\rule{0.967pt}{0.116pt}}
\multiput(1153.00,652.17)(10.994,6.000){2}{\rule{0.483pt}{0.400pt}}
\multiput(1166.00,659.59)(0.874,0.485){11}{\rule{0.786pt}{0.117pt}}
\multiput(1166.00,658.17)(10.369,7.000){2}{\rule{0.393pt}{0.400pt}}
\multiput(1178.00,666.59)(0.874,0.485){11}{\rule{0.786pt}{0.117pt}}
\multiput(1178.00,665.17)(10.369,7.000){2}{\rule{0.393pt}{0.400pt}}
\multiput(1190.00,673.59)(1.123,0.482){9}{\rule{0.967pt}{0.116pt}}
\multiput(1190.00,672.17)(10.994,6.000){2}{\rule{0.483pt}{0.400pt}}
\multiput(1203.00,679.59)(0.874,0.485){11}{\rule{0.786pt}{0.117pt}}
\multiput(1203.00,678.17)(10.369,7.000){2}{\rule{0.393pt}{0.400pt}}
\multiput(1215.00,686.59)(1.033,0.482){9}{\rule{0.900pt}{0.116pt}}
\multiput(1215.00,685.17)(10.132,6.000){2}{\rule{0.450pt}{0.400pt}}
\multiput(1227.00,692.59)(0.874,0.485){11}{\rule{0.786pt}{0.117pt}}
\multiput(1227.00,691.17)(10.369,7.000){2}{\rule{0.393pt}{0.400pt}}
\multiput(1239.00,699.59)(1.123,0.482){9}{\rule{0.967pt}{0.116pt}}
\multiput(1239.00,698.17)(10.994,6.000){2}{\rule{0.483pt}{0.400pt}}
\multiput(1252.00,705.59)(0.874,0.485){11}{\rule{0.786pt}{0.117pt}}
\multiput(1252.00,704.17)(10.369,7.000){2}{\rule{0.393pt}{0.400pt}}
\multiput(1264.00,712.59)(1.033,0.482){9}{\rule{0.900pt}{0.116pt}}
\multiput(1264.00,711.17)(10.132,6.000){2}{\rule{0.450pt}{0.400pt}}
\multiput(1276.00,718.59)(0.950,0.485){11}{\rule{0.843pt}{0.117pt}}
\multiput(1276.00,717.17)(11.251,7.000){2}{\rule{0.421pt}{0.400pt}}
\multiput(1289.00,725.59)(1.033,0.482){9}{\rule{0.900pt}{0.116pt}}
\multiput(1289.00,724.17)(10.132,6.000){2}{\rule{0.450pt}{0.400pt}}
\multiput(1301.00,731.59)(0.874,0.485){11}{\rule{0.786pt}{0.117pt}}
\multiput(1301.00,730.17)(10.369,7.000){2}{\rule{0.393pt}{0.400pt}}
\multiput(1313.00,738.59)(1.033,0.482){9}{\rule{0.900pt}{0.116pt}}
\multiput(1313.00,737.17)(10.132,6.000){2}{\rule{0.450pt}{0.400pt}}
\multiput(1325.00,744.59)(0.950,0.485){11}{\rule{0.843pt}{0.117pt}}
\multiput(1325.00,743.17)(11.251,7.000){2}{\rule{0.421pt}{0.400pt}}
\multiput(1338.00,751.59)(1.033,0.482){9}{\rule{0.900pt}{0.116pt}}
\multiput(1338.00,750.17)(10.132,6.000){2}{\rule{0.450pt}{0.400pt}}
\multiput(1350.00,757.59)(0.874,0.485){11}{\rule{0.786pt}{0.117pt}}
\multiput(1350.00,756.17)(10.369,7.000){2}{\rule{0.393pt}{0.400pt}}
\multiput(1362.00,764.59)(1.123,0.482){9}{\rule{0.967pt}{0.116pt}}
\multiput(1362.00,763.17)(10.994,6.000){2}{\rule{0.483pt}{0.400pt}}
\multiput(1375.00,770.59)(0.874,0.485){11}{\rule{0.786pt}{0.117pt}}
\multiput(1375.00,769.17)(10.369,7.000){2}{\rule{0.393pt}{0.400pt}}
\multiput(1387.00,777.59)(0.874,0.485){11}{\rule{0.786pt}{0.117pt}}
\multiput(1387.00,776.17)(10.369,7.000){2}{\rule{0.393pt}{0.400pt}}
\multiput(1399.00,784.59)(1.033,0.482){9}{\rule{0.900pt}{0.116pt}}
\multiput(1399.00,783.17)(10.132,6.000){2}{\rule{0.450pt}{0.400pt}}
\multiput(1411.00,790.59)(0.950,0.485){11}{\rule{0.843pt}{0.117pt}}
\multiput(1411.00,789.17)(11.251,7.000){2}{\rule{0.421pt}{0.400pt}}
\multiput(1424.00,797.59)(1.033,0.482){9}{\rule{0.900pt}{0.116pt}}
\multiput(1424.00,796.17)(10.132,6.000){2}{\rule{0.450pt}{0.400pt}}
\end{picture}
\end{center}
\end{figure}

\clearpage
\begin{thebibliography}{99}
\bibitem{Domb58}
Domb C.,
{\it Nuovo Ciment}, {\bf 9} Suppl. (1958) 9.

\bibitem{Temperley59}
Temperley H.N.V.,
{\it Proc. Phys. Soc.}, {\bf 74} (1959) 183.

\bibitem{Burley60}
Burley D.M.,
{\it Proc. Phys. Soc.}, {\bf 75} (1960) 262.

\bibitem{GauntFisher65}
Gaunt D.S. and Fisher M.E.,
{\it J. Chem. Phys.}, {\bf 43} (1965) 2840.

\bibitem{Runnels65}
Runnels L.K.,
{\it Phys. Rev. Lett.}, {\bf 15} (1965) 581.

\bibitem{RunnelsCombs66}
Runnels L.K. and Combs L.L.,
{\it J. Chem. Phys.}, {\bf 45} (1966) 2482.

\bibitem{ReeChesnut66}
Ree F.H. and Chesnut D.A.,
{\it J. Chem. Phys.}, {\bf 45} (1966) 3983.

\bibitem{Baxteretal80}
Baxter R.J., Enting I.G., and Tsang S.K.,
{\it J. Stat. Phys.}, {\bf 22} (1980) 465.

\bibitem{Racz80}
R\`acz Z.,
{\it Phys. Rev. B}, {\bf 21} (1980) 4012.

\bibitem{WoodGoldfinch80}
Wood D.W. and Goldfinch M.,
{\it J. Phys. A}, {\bf 13} (1980) 2781.

\bibitem{Meirovitch83}
Meirovitch H.,
{\it J. Stat. Phys.}, {\bf 30} (1983) 681.

\bibitem{KamieniarzBlote93a}
Kamieniarz G. and Bl\"ote H.W.J.,
{\it J. Phys. A}, {\bf 26} (1993) 201.

\bibitem{KamieniarzBlote93b}
Kamieniarz G. and Bl\"ote H.W.J.,
{\it J. Phys. A}, {\bf 26} (1993) 6679.

\bibitem{Gaunt67}
Gaunt D.S.,
{\it J. Chem. Phys.}, {\bf 46} (1967) 3237.

\bibitem{Baxter80}
Baxter R.J.,
{\it J. Phys. A}, {\bf 13} (1980) L61.

\bibitem{BaxterPearce82}
Baxter R.J. and Pearce P.A.,
{\it J. Phys. A}, {\bf 15} (1982) 897.

\bibitem{Runnelsetal67}
Runnels L.K., Combs L.L., and Salvant J.P.,
{\it J. Chem. Phys.}, {\bf 47} (1967) 4015.

\bibitem{Yamagata95a}
Yamagata A.,
{\it Physica A}, {\bf 215} (1995) 511.

\bibitem{Yamagata95b}
Yamagata A.,
{\it Physica A}, {\bf 222} (1995) 119.

\bibitem{Yamagata95c}
Yamagata A.,
{\it Physica A} in print.

\bibitem{FerrenbergLandau91}
Ferrenberg A.M. and Landau D.P.,
{\it Phys. Rev. B}, {\bf 16} (1991) 5081.

\bibitem{Binder79}
Binder K.,
in {\it Monte Carlo Methods in Statistical Physics},
edited by K.~Binder
(Springer, Berlin) 1979 p.~1.

\bibitem{BinderStauffer87}
Binder K. and Stauffer D.,
in {\it Applications of the Monte Carlo Method in Statistical Physics},
2nd ed.,
edited by K. Binder
(Springer, Berlin) 1987 p.~1.

\bibitem{ItoKanada88}
Ito N. and Kanada  Y.,
{\it Supercomputer}, {\bf 5} (1988) 31.

\bibitem{ItoKanada90}
Ito N. and Kanada Y.,
{\it Proceedings of Supercomputing '90}
(IEEE Computer Society Press, Los Alamitos) 1990 p.~753.

\bibitem{Fisher70}
Fisher M.E.,
in {\it Critical Phenomena, Proc. 1970 Enrico Fermi Summer School},
Vol. 51,
edited by M.S. Green
(Academic Press, New York) 1970 p.~1.

\bibitem{Barber83}
Barber M.N.,
in {\it Phase Transitions and Critical Phenomena}, Vol. 8,
edited by C. Domb and J.L. Lebowitz
(Academic Press, London) 1983 p.~145.

\bibitem{Privman90}
Privman V.,
in {\it Finite Size Scaling and Numerical Simulation of Statistical Systems},
edited by V. Privman
(World Scientific, Singapore) 1990 p.~1.

\bibitem{Pressetal92}
Press W.H., Teukolsky S.A., Vetterling W.T., and Flannery B.P.,
{\it Numerical Recipes in C: The Art of Scientific Computing}, 2nd ed.
(Cambridge University Press, Cambridge) 1992 Chapters 6 and 15.
\end{thebibliography}
\end{document}